\documentclass[12pt]{article}
\pdfoutput=1
\usepackage{amsmath,amssymb,graphicx} 
\usepackage{epsf}
\usepackage{pstricks}
\usepackage{cite}

\newcommand{\beq}{\begin{eqnarray}}
\newcommand{\eeq}{\end{eqnarray}}

\newcommand{\drawsquare}[2]{\hbox{%
\rule{#2pt}{#1pt}\hskip-#2pt
\rule{#1pt}{#2pt}\hskip-#1pt
\rule[#1pt]{#1pt}{#2pt}}\rule[#1pt]{#2pt}{#2pt}\hskip-#2pt
\rule{#2pt}{#1pt}}

\newcommand{\fund}{\drawsquare{6.5}{0.4}}

\newcommand{\asymm}{\raisebox{-3.5pt}{\drawsquare{6.5}{0.4}\hskip-6.9pt%
        \raisebox{6.5pt}{\drawsquare{6.5}{0.4}}}}

\newcommand{\centeron}[2]{{\setbox0=\hbox{#1}\setbox1=\hbox{#2}\ifdim
\wd1>\wd0\kern.5\wd1\kern-.5\wd0\fi
\copy0

\kern-.5\wd0\kern-.5\wd1\copy1\ifdim\wd0>\wd1
                                       \kern.5\wd0\kern-.5\wd1\fi}}
\newcommand{\ltap}{\>\centeron{\raise.35ex\hbox{$<$}}
                               {\lower.65ex\hbox{$\sim$}}\>}
\newcommand{\gtap}{\>\centeron{\raise.35ex\hbox{$>$}}
                               {\lower.65ex\hbox{$\sim$}}\>}

\newcommand\ZZ{\hbox{\zfont Z\kern-.4emZ}}
\font\zfont = cmss10 

\begin{document}
\begin{titlepage}
\begin{flushright}
\end{flushright}

\vskip.5cm
\begin{center}
{\huge \bf
Electroweak Symmetry Breaking \\
\vskip0.2cm
From Monopole Condensation
}

\vskip.1cm
\end{center}
\vskip0.2cm

\begin{center}
{\large \bf Csaba Cs\'aki$^{a}$,
Yuri Shirman$^{b}$, {\rm and}
John Terning$^{c,d}$}
\end{center}
\vskip 8pt

\begin{center}
$^{a}$ {\it Institute for High Energy Phenomenology\\
Newman Laboratory of Elementary Particle Physics\\
Cornell University, Ithaca, NY 14853, USA } \\
$^{b}$ {\it
Department of Physics, University of California, Irvine, CA
92697.} \\

$^{c}$ {\it
Department of Physics, University of California, Davis, CA
95616.} \\

$^{d}$ {\it
CERN, Physics Department, Theory Unit, Geneva, Switzerland.}\\

\vspace*{0.3cm}
{\tt   csaki@cornell.edu, yshirman@uci.edu, terning@physics.ucdavis.edu}
\end{center}

\vspace*{0.2cm}

\vglue 0.3truecm

\begin{abstract}
\vskip 3pt \noindent We examine models where massless chiral fermions with both ``electric" and ``magnetic"
hypercharges could form condensates.  When some of the fermions are also electroweak doublets
such condensates can break the electroweak gauge symmetry down to electromagnetism in the
correct way. Since ordinary hypercharge is weakly coupled at the TeV scale, magnetic hypercharge  is strongly coupled and can potentially drive  the condensation.  Such models are similar to technicolor, but with hypercharge playing the role of the technicolor gauge group, so the standard model gauge group breaks itself. A heavy top mass can be generated via the Rubakov-Callan effect and could thus decouple the scale of flavor physics from the electroweak scale.

\end{abstract}

\end{titlepage}

\newpage


\section{Introduction}
\label{sec:intro}
\setcounter{equation}{0}
\setcounter{footnote}{0}
All currently  viable models of electroweak symmetry breaking are fine tuned at some level,   while theories of electroweak symmetry breaking that are not fine-tuned, like technicolor, are not viable, due to difficulties with electroweak precision observables (EWPO's) and flavor changing neutral currents (FCNC's).  In this paper we explore a model of monopole condensation which is not fine tuned and could potentially avoid some of the difficulties of technicolor models. It is known from the work of Argyres and Douglas \cite{ArgyresDouglas} that there are consistent field theories with  both magnetic and electric massless charged particles.   These ${\mathcal N}=2$  supersymmetric  theories are expected to reach IR fixed points. However in more general theories of massless magnetic and electric charges it may be possible  that one type of charge outweighs the other in the running of the coupling and that a fixed point cannot be reached. In this case, either the electric or magnetic coupling is driven into the strongly coupled regime. We then might expect to enter either a Higgs  phase or a confining phase. Due to  electromagnetic duality these can be the same thing,  since an electrically charged Higgs condensate confines magnetic charges and vice versa \cite{tHooftMandelstam}.   Here we would like to explore the possibility that a bilinear condensate of magnetically charged fermions breaks the electroweak  gauge group down to electromagnetism. An even more exciting possibility is that these new particles have magnetic hypercharges so that the standard model gauge group breaks  itself.
The heavy top mass could be explained via generation of a Rubakov-Callan operator, thereby decoupling the scale of flavor physics from the of electroweak symmetry breaking.

\section{Monopoles and Dyons}
\label{sec:MonopolesDyons}
\setcounter{equation}{0}
\setcounter{footnote}{0}

Magnetic monopoles have been a source of fascination for a very long time~\cite{reviews}. J.J. Thomson \cite{Thomson} calculated in 1904 the angular momentum of the electromagnetic (EM) field in the presence of an electric charge $q$ (in units of $e$) and a magnetic charge $g$ (in units of $4\pi/e$). He found\footnote{We use  Heavyside-Lorentz units where the Coulomb law is $\vec{E}= \frac{e\,q }{4\pi r^2} {\hat r}$, and $c=\hbar =1$.}
  \beq
 \vec{J}=q g\, \hat{r}
 \label{angularmom}
 \eeq
where $ \hat{r}$ is a unit vector pointing from the charge to the  magnetic  monopole.

Later Dirac \cite{Dirac} showed that in a quantum mechanical theory a monopole could be thought of as a gauge configuration with an unobservable singular string, with the result that $q g$  is quantized in units of  half  integers. This result also quantizes the angular momentum in (\ref{angularmom}) as one would expect.
Dirac was also able to write down a Lagrangian \cite{DiracLagrangian}  for the interaction of electric and magnetic charges.
A new  non-local contribution to the field strength tensor was required to account for the electromagnetic interactions of the monopole via the Dirac string.

 Schwinger extended the idea of monopoles to include dyons  \cite{Schwinger:1969ib} which have both electric and magnetic charge.  Two dyons must satisfy a generalized charge quantization condition \cite{Schwinger}:
\beq
q_1\,g_2-q_2\,g_1= \frac{n}{2}~,
\label{quantization}
\eeq
where $n$ is an integer. Later, Zwanziger~\cite{Zwanziger}  was able to rewrite Dirac's Lagrangian  in a local but non-Lorentz invariant  form.
Zwanziger's Lagrangian contains two gauge fields: one that couples to electric charges and and one that couples to  magnetic charges, however the form of the kinetic terms is such that there are only two propagating degrees of freedom. This description is very useful when considering explicit calculations of quantum effects involving monopoles.

In the 1970's `t Hooft and Polyakov \cite{tHooftPolyakov}  found that monopoles could be realized as topological solitons of a broken Yang-Mills theory and that such solitons were a necessary feature of any grand unified theory (GUT). `t Hooft and Mandelstam also explained that just as an electrically charged condensate confines magnetic charges in a superconductor, a magnetically charged condensate confines electric charges \cite{tHooftMandelstam}.

Another fascinating development regarding the unusual properties of monopoles happened in the early 80's, when Rubakov and Callan \cite{RubakovCallan} showed that in addition to electromagnetic interactions, monopoles and charges must have other nontrivial interactions in order to maintain the consistency of the  theory. Consider low energy s-wave scattering of a charge off a monopole, and recall that  the angular momentum of the electromagnetic field points from the  charge towards the monopole as in (\ref{angularmom}). In a head-on collision, there are no electromagnetic forces on the particles but simple forward scattering would result in a final state where the angular momentum vector has flipped sign, which would not conserve angular momentum.  There must be new unsuppressed contact interactions in order to unitarize low-energy s-wave scattering, which one might not expect from a low-energy effective field theory approach. Applying this
phenomenon  to GUTs  they found that protons scattering off monopoles would catalyze proton decay at QCD rates, even though there is not enough energy to produce a broken GUT gauge boson.

In the 1990s Seiberg and Witten showed that ${\mathcal N}=2$  supersymmetric gauge theories have points on the moduli space with massless monopoles or dyons \cite{SeibergWitten}. Since these theories are supersymmetric, there are both bosonic and fermionic massless monopoles. Argyres and Douglas~\cite{ArgyresDouglas} then showed that in more elaborate theories the separate points with massless monopoles and dyons can be moved on top of each other so that both types of massless charges exist in the same low-energy theory. Recently we showed~\cite{cst} how to calculate anomalies for coexisting massless fermionic monopoles and dyons using the methods of~\cite{Zwanziger,ArgyresDouglas}.

While many exciting and unusual features of monopoles have been identified over the past century, there hasn't been much effort devoted to finding potential uses of monopoles for particle physics phenomenology (with the notable exception of Schwinger's attempt to explain the origin of strong interactions via monopoles~\cite{Schwinger}). Here we initiate the study of possible uses of monopoles for applications to the physics of the TeV scale and electroweak symmetry breaking.

\section{A Toy Model}
\label{sec:Toymodel}
\setcounter{equation}{0}
\setcounter{footnote}{0}

Our aim is to find a matter content of monopoles and/or dyons that could potentially give rise to realistic natural electroweak symmetry breaking. For this the spectrum must satisfy the following properties:

\begin{itemize}
\item to have strong dynamics in the IR the monopoles/dyons should be massless,

\item to avoid the hierarchy problem, the monopoles/dyons should be fermionic,

\item all anomalies~\cite{cst} including the mixed electric and magnetic anomalies should cancel,

\item an SU(2)$_R$ custodial symmetry should protect the $T$-parameter,

\item to avoid confinement of the electric charges, the magnetic charges should be vector-like, so that the condensates can be magnetically neutral,

\item the Dirac-Schwinger quantization condition (\ref{quantization}) should be satisfied.
\end{itemize}

To satisfy these conditions it is sufficient to add to the SM one generation of fields which, in addition to the usual quantum numbers, also carry magnetic hypercharge.
These fields would be analogous to the techniquarks, whose dynamics gives rise to EWSB.
None of the SM fields carry magnetic charges of course.
\beq
\begin{array}{c|ccrr}
& SU(3)_c & SU(2)_L & U(1)_Y^{el} & U(1)_Y^{mag} \\
\hline
Q & \fund & \fund & \frac{1}{6} & 3    \vphantom{\raisebox{3pt}{\asymm}}\\
L & 1 & \fund & -\frac{1}{2} & -9  \vphantom{\raisebox{3pt}{\asymm}}\\
{\bar U} & {\bar \fund} & 1 & -\frac{2}{3} & -3     \vphantom{\raisebox{3pt}{\asymm}}\\
{\bar D} & {\bar \fund} & 1& \frac{1}{3} & -3      \vphantom{\raisebox{3pt}{\asymm}}\\
{\bar N} & 1 & 1 & 0 & 9    \vphantom{\raisebox{3pt}{\asymm}}\\
{\bar E} & 1& 1 &1 & 9     \vphantom{\raisebox{3pt}{\asymm}}\\
\end{array}
\label{toymodel}
\eeq
It is quite obvious that with these charges all anomaly conditions, including the ones involving the magnetic charges~\cite{cst} of the sort $\sum_j {\rm Tr}\,T^a_{r_j} T^b_{r_j} g_j=0, \ \sum_j g_{j}^2 q_j=0, \ldots$, are satisfied. The reason is that the magnetic charges have been chosen to be proportional to the $B-L$ charges of the SM fields, and thus even if one treated $U(1)_Y^{mag}$ as a separate U(1) all anomalies would cancel, implying the cancelation of all mixed magnetic anomalies as well.

The overall normalization of the magnetic hypercharge has been chosen such that the Dirac-Schwinger quantization condition (\ref{quantization}) is satisfied for any pair of fields.  Custodial symmetry is obtained by making the magnetic hypercharges of the right handed singlets equal to each other. From the point of view of the strong magnetic hypercharge interactions there is an SU(2)$_L\times$SU(2)$_R$ symmetry, which is weakly broken by electric hypercharge.

There are no tools to directly analyze the IR properties of this theory, except perhaps lattice simulations. There are two plausible low-energy phases of this model. One possibility would be for it to sit at an IR fixed point, similar to those of Argyres and Douglas~\cite{ArgyresDouglas}. In this phase the theory would not be useful for electroweak symmetry breaking. The other plausible option is that the full non-perturbative $\beta$-function of the theory is very different from the naive one-loop $\beta$-function, and that  the electric hypercharge from 3+1 generations actually dominates over the contributions of the magnetic hypercharge from 1 generation. In this case the electric hypercharge would become weaker as one goes towards the IR, while the magnetic hypercharge would keep increasing (and by our hypothesis its contribution to the $\beta$-function would keep decreasing). In such a scenario the theory is driven to a very strongly interacting magnetic theory, and magnetic charges could condense as quarks do in QCD. Such chiral symmetry breaking is observed in strongly coupled $U(1)$ theories on the lattice \cite{lattice1,lattice2}. The charges in (\ref{toymodel}) have been chosen such that the plausible set of condensates have the right quantum numbers to play the role of the SM Higgs:
\begin{eqnarray}
&& Q \bar{D}  \sim (1,2, \frac{1}{2}) \sim H , \ \
Q \bar{U} \sim (1,2, -\frac{1}{2}) \sim H^* , \nonumber \\
&& L \bar{E} \sim (1,2, \frac{1}{2}) \sim H , \ \
L \bar{N}  \sim (1,2, -\frac{1}{2}) \sim H^* .
\label{quantum numbers}
\end{eqnarray}
Thus we need to assume that the upper component of the doublet $Q$, aka $U_L$,  condenses with $\bar{U}$, the lower component of $Q$, aka $D_L$  with $\bar{D}$, the upper component of $L$, aka $N_L$, with $\bar{N}$ and the lower component of $L$, aka $E_L$, with $\bar{E}$:
\begin{equation}
\langle U_L \bar{U} \rangle \sim \langle D_L \bar{D} \rangle \sim \langle N_L \bar{N} \rangle \sim \langle E_L \bar{E} \rangle \sim \Lambda_{mag}^d
\label{condensates}
\end{equation}
where $\Lambda_{mag}$ is the scale of condensation that would be dynamically generated by the strong magnetic hypercharge interactions, and $d$ is the a priori unknown scaling dimension of the bilinear operators. In the rest of the paper we will assume that the low-energy dynamics of the theory is indeed of this type: magnetic interactions generate a mass gap of order $\Lambda_{mag}$, all particles carrying magnetic hypercharge pick up a dynamical mass of this order, and the condensates in (\ref{condensates}) are formed giving rise to electroweak symmetry breaking as in ordinary QCD and technicolor theories.

We should check whether the conjecture above agrees with our experience in QCD-like theories.
The Dirac quantization condition
\beq
\left( \frac{1}{6}\right)^2\alpha_Y 3^2\alpha_m= \frac{1}{4}
\eeq
and the  hypercharge coupling $\alpha_Y \sim 0.0102$ lead to $\alpha_m\sim 98$
while one would naively expect condensation to happen for $\alpha_m \sim 4\pi$. However we do not have any experience with theories containing massless electric and magnetic charges and such theories have not been studied in lattice simulations. If lattice simulations were to confirm the naive expectation, one can still use the mechanism outlined above for electroweak symmetry breaking, except that one would need to use a U(1) different from hypercharge, for which the coupling constant can be freely adjusted.
Later in the paper we will develop a more realistic model that has a much smaller value for $\alpha_m$.

\section{Review of Rubakov-Callan Operators}
\label{sec:CallanRubakov}
\setcounter{equation}{0}
\setcounter{footnote}{0}

As mentioned earlier, technicolor theories suffer from two main issues: a large correction to the S-parameter and the difficulty of generating sufficiently large top mass without also generating large FCNC's. The S-parameter in our theory is incalculable, one may however hope that its magnitude is reduced to something comparable to a fourth generation model, since the size of the strongly interacting gauge group is minimal, a U(1). On the other hand, the Rubakov-Callan effect may provide a new avenue for generating a large fermion mass. In this section we review the essence of the Rubakov-Callan effect, and show that in the toy model presented above such effects do not lead to a large top mass. In section \ref{sec:model} we present a modified model where the Rubakov-Callan effect indeed gives rise to a large top mass.

\begin{figure}
\begin{center}
\includegraphics[width=0.6\hsize]{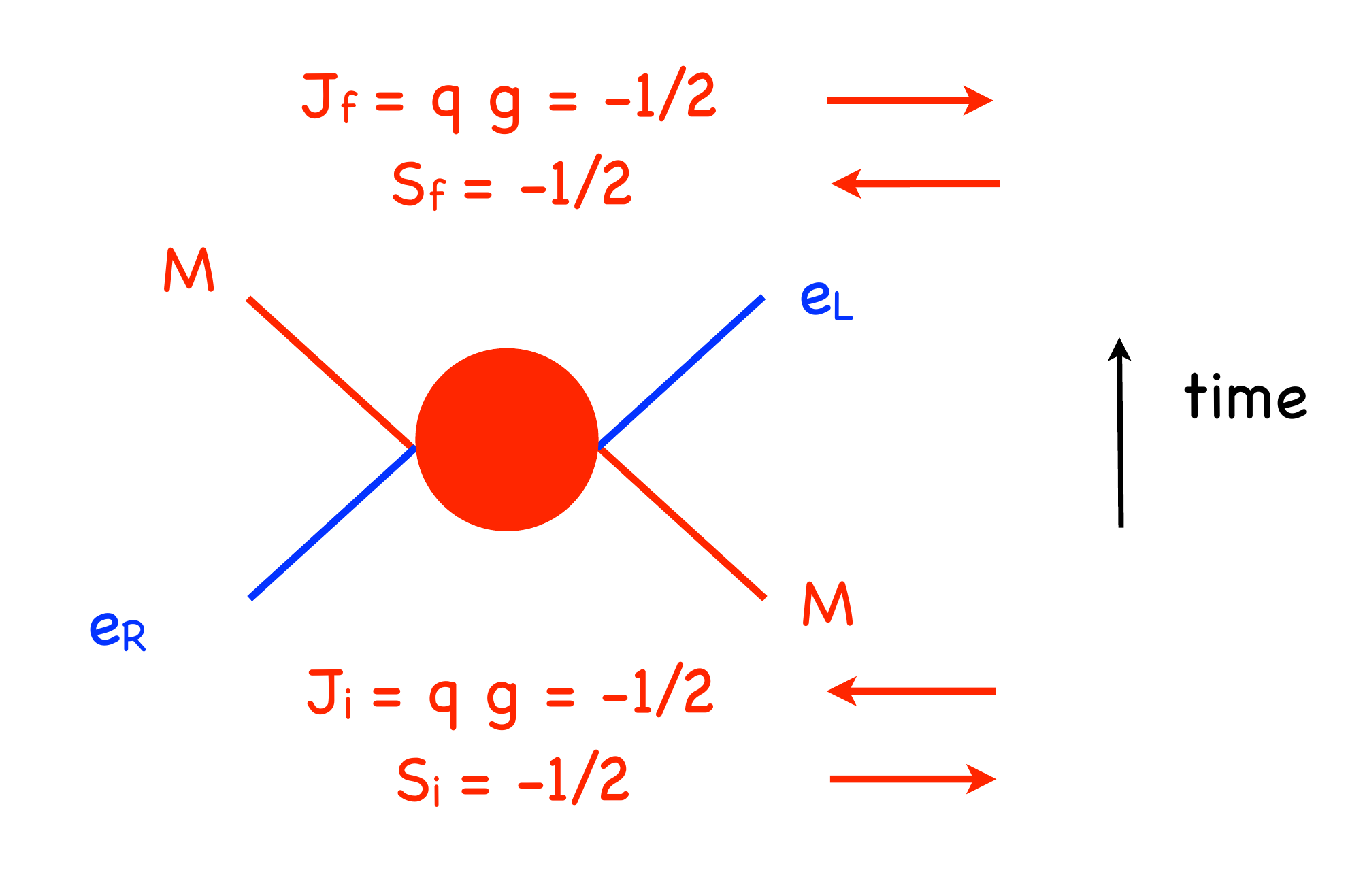}
\caption{\label{fig:RC} The scattering of electrons on scalar monopoles leading to the Rubakov-Callan effect, that implies the presence of chirality flipping marginal operators.}
\end{center}
\end{figure}

Rubakov and Callan considered head-on scattering of electrons and scalar monopoles (see Fig.~\ref{fig:RC}). Assume that we start out with a LH electron. The angular momentum of the EM field is 1/2, and opposite to the angular momentum of the electron, such that the total angular momentum in the initial state is zero. Once the electron passes through the monopole, the direction of the angular momentum in the EM field flips, since it always points from the charge to the monopole. In order to conserve angular momentum, either the chirality or the electric charge of the electron has to flip. If charge is conserved, then new chirality violating operators of the form
\begin{equation}
e_L  \bar{e} M^* M
\end{equation}
must be present, which are not suppressed by a high scale (that is this operator is marginal).

Applying this  idea to EWSB, we would like to find such Rubakov-Callan operators for fermionic monopoles, i.e. 4-fermi operators that lead to the top mass after monopole condensation. However, it is easy to see that in the model of Sec.~\ref{sec:Toymodel} this does not actually happen. An example of the type of operator one would like is
\beq
t_L \sigma^{\mu\nu}  \bar{t}\, U_L^\dagger \bar{ \sigma}_{\mu\nu} \bar{U}^\dagger~.
\eeq
which, after Fierz rearrangement contains the operator
\begin{equation}
t_L \bar{t} \left(U_L \bar{U}\right)^\dagger~.
\label{topmass}
\end{equation}
This operator is contained in the gauge invariant
\beq
\lambda^{(u)}_{ij} q^i_L \bar{u}^j \left(Q_L \bar{U}\right)^\dagger~,
\label{generaltopmass}
\eeq
where the indices $i$ and $j$ refer to generation number.
 For this operator to exist one would need angular momentum conservation in the scattering $t_R U_L \to t_L U_R$. However, the angular momentum of the EM field in the initial state is $3\times \frac{2}{3}=2$, while in the final state $-3 \times \frac{1}{6}=-\frac{1}{2}$. A half-integer difference in the angular momentum can not be made up with any combination of chirality flips, so (\ref{topmass}) is not one of the Rubakov-Callan operators induced here.

\section{Non-Abelian Magnetic Charges}
\label{sec:Z_N}
\setcounter{equation}{0}
\setcounter{footnote}{0}

The operator (\ref{topmass}) was not induced in our toy model because the Dirac quantization condition required a large values of magnetic charges. This further lead to large angular momenta in the electromagnetic field in the scattering process, and the incoming wave could not be in an s-wave. This situation is in fact similar to the issue extensively studied in the early 80's, when the question of whether a monopole with the minimal Dirac charge $\tilde g_D = \frac{1}{2} \frac{4\pi}{e}$, i.e. $g_D=  \frac{1}{2} $, could actually co-exist with fractionally charged quarks. Naively one would think that the Dirac quantization condition in the presence of a down quark would imply that the minimal magnetic charge is $3\,g_D$. However, this is only true if the magnetic generator is completely aligned with electromagnetism.  It was realized by 't Hooft that the minimal Dirac charge is indeed allowed, but only if monopoles with charges smaller than $3\, g_D$ also carry a color magnetic charge, which will provide an additional contribution to the Dirac quantization condition \cite{nonabelianmagnetic,reviews}.

This teaches us that the magnetic hypercharge quantum numbers may be lowered with a different embedding of the magnetic $U(1)$ group in the SM. Until now we assumed that the magnetic $U(1)$ is completely aligned with the magnetic hypercharge. However it could as well be a combination of some of the magnetic non-Abelian generators with hypercharge.  During the study of $SU(5)$ GUT monopoles \cite{nonabelianmagnetic} it was found that precisely this happens in the standard $SU(5) \to SU(3)_c\times SU(2)\times U(1)_Y$ breaking. Due to the non-trivial embedding of magnetic charge into $SU(5)$ certain color, weak,  and hypercharge group elements get identified, and the actual global structure of the gauge group is $(SU(3)_c \times SU(2)_L\times U(1)_Y)/Z_6$. This leads to an entanglement of the magnetic generators, leading to a minimally charged Dirac monopole, which however also carries magnetic charge under color and weak interactions. To describe such monopoles,
we can first pick a gauge  where the long range fields point in a particular direction in the gauge space.  Thus for a monopole with charge $g$  we can have the following $SU(3)_c \times SU(2)_L\times U(1)_Y$ long range fields \cite{nonabelianmagnetic}:
\beq
{\vec B}_Y^a &=&  \frac{g}{g_Y} \frac{\hat r} { r^2}~, \label{nonab1} \\
{\vec B}_L^a &=& \delta_L^{a3}\,  \frac{g \, \beta_L}{g_L} \, \frac{\hat r}{ r^2}~, \label{nonab2} \\
{\vec B}_c^a &=&  \delta_c^{a8}\,  \frac{g \,  \beta_c }{g_c}\, \frac{ {\hat r}}{r^2}~, \label{nonab3}
\label{longrange}
\eeq
where, at this point, $\beta_L$ and $\beta_c$ are simply parameters that fix the relative strength of magnetic couplings under different gauge groups. In this theory the usual Dirac  calculation of transporting a charge around the monopole
\beq
\int_{loop} e \, q \, A^\mu dx_\mu
\eeq
is generalized to
\beq
\int_{loop} \left( g_c \,T_c^a G^{a \mu} +g_L\, T_L^a\, W^{a\mu}+g_Y Y B^\mu \right)dx_\mu~,
\label{loop}
\eeq
which now can be thought of as a (diagonal) matrix, and all of the eigenvalues have to satisfy the Dirac quantization condition. The matrix structure is obvious in a GUT theory and, more generally, can be easily determined from quantum numbers of fields in (\ref{loop}). One can evaluate (\ref{loop}) in the standard way: the vector potential for a Dirac string at $\theta =-\pi$ giving rise to (\ref{nonab1})-(\ref{nonab3}) in a suitable gauge is
\beq
{\vec A}_Y &=& \frac{g}{g_Y}  \, \frac{ 1-\cos \theta}{ r \sin \theta}\,{\hat e}_\phi ~. \label{nonabA1} \\
{\vec A}_L^a &=& \delta_L^{a3}\,  \frac{g \, \beta_L}{g_L}   \, \frac{ 1-\cos \theta}{r \sin \theta}\,{\hat e}_\phi ~, \label{nonabA2} \\
{\vec A}_c^a &=&   \delta_c^{a8}\, \frac{g \,  \beta_c }{g_c} \,  \frac{ 1-\cos \theta}{ r \sin \theta}\,{\hat e}_\phi ~, \label{nonabA3}
\label{longrangeA}
\eeq
Then the absence of an observable Aharonov-Bohm phase of a particle which carries color, SU(2)$_L$ and U(1)$_Y$ charges leads to the generalized Dirac quantization condition:
\beq
4\pi \left(  T_c^8 \, g \, \beta_c  + \,T_L^3\,  g\, \beta_L    + Y g \right) = 2 \pi n~.
\label{Diracnonab}
\eeq
This condition has to be satisfied for  all components of the diagonal matrix and all pairs of fields in the model.

In general, the choice of $\beta_L$ (and $\beta_c$) satisfying (\ref{Diracnonab}) is not unique.
The choice of $\beta_L$ is fixed by the requirement that monopoles do not couple to the $Z$-boson --- otherwise it would be hard to imagine how electroweak precision corrections could possibly be suppressed, and there would also be modifications to the $Z$ width \cite{DeRujula}.
Furthermore if part of the magnetic charge of the monopoles was carried by $Z$, the 't Hooft-Mandelstam argument would imply that the monopoles are confined once $Z$ obtains mass.
Monopoles remain unconfined if the $Z$ does not couple magnetically to the monopoles. This can be ensured if the direction of the magnetic fields in (\ref{nonab1})-(\ref{nonab3}) is such that only the vector potential $A^\mu$ corresponding to the photon (and possibly some gluons as well) is turned on, that is, there are only massless gauge bosons in the long range fields of the monopoles. With the usual embedding of electric charge into the electroweak group,
\beq
e A^\mu = g_L A^{3\mu}_L +g_Y A^\mu_Y~,
\eeq
  one can see that the choice $\beta_L=1$ will ensure that only the photon, and not the $Z$, couples to the magnetic charge of  the monopoles.  Using eq. (\ref{nonabA1}) and (\ref{nonabA2}) we can see that with $\beta_L=1$ the electroweak part of the long range field is just
\beq
{\vec A} &=& \frac{g}{e}  \, \frac{ 1-\cos \theta}{ r \sin \theta}\,{\hat e}_\phi ~, \label{emfield}
 \eeq
which corresponds to a magnetic charge for the monopole  $\tilde g= g \, 4 \pi/e$  under $U(1)_{em}$.  It is also clear that  for monopoles with $\beta_L=1$ there is no difference in calculating the Dirac quantization conditions above or below the electroweak scale because
\beq
q  = (T_L^3 +Y) ~,
\eeq
and (\ref{Diracnonab}) can also be written as:
\beq
  T_c^8 \, g \, \beta_c  + q \, g = \frac{n}{2}~.
\eeq
  For dyons (as we are considering) the Dirac-Schwinger condition becomes
\beq
 (q_1 \, g_2-q_2 \, g_1)+  (T^8_{c1} \, g_2 \, \beta_{c2}- T^8_{c2} \, g_1\, \beta_{c1}) = \frac{n}{2}~.
\eeq
Here group generators $T^8$ correspond to electric (but not magnetic) color charges of the SM fields and dyons.
Indeed, a very peculiar feature of non-Abelian magnetic charges is that such states do not appear as  full representations of the unbroken gauge group. In the presence of non-Abelian magnetic charges,  global color rotations are not defined unless they leave the long-range gauge fields (\ref{longrangeA}) invariant \cite{globalcolor}. Thus the non-Abelian color magnetic charge is completely described by a single number $g \beta_c/g_c \, $ rather than the matrix of charges, $g_c T^a$, associated with a color triplet.

\section{A Model with a heavy top}
\label{sec:model}
\setcounter{equation}{0}
\setcounter{footnote}{0}

We now modify the toy model of section~\ref{sec:Toymodel} by assuming that the global structure of the gauge group is $SU(3)_c \times SU(2)_L\times U(1)_Y/Z_6$,  and just like the ordinary $SU(5)$ GUT that the monopoles carry non-Abelian magnetic charges for both the color and electroweak groups. We will take $\beta_c\ne 0$ for the quark-like monopoles, and $\beta_c=0$ for the lepton-like monopoles. As explained in the previous section, we will assume that $\beta_L=1$ for every field carrying monopole charges, which as we saw is equivalent to the assumption that the $SU(2)_L\times U(1)_Y$ magnetic charge points in the direction of the ordinary photon. In this case (as we have seen above) the effective magnetic charge is just the same as the magnetic hypercharge
\beq
g_{em}=Y^{mag}=g~.
\eeq
The field content is given by
 \beq
\begin{array}{c|cccc}
& SU(3)_c & SU(2)_L& U(1)_Y^{el} & U(1)_Y^{mag} \\
\hline
Q_L & \fund^m & \fund^m & \frac{1}{6} &\frac{1}{2}   \vphantom{\raisebox{4pt}{\asymm}}\\
L_L & 1 &\fund^m  & -\frac{1}{2} & -\frac{3}{2}   \vphantom{\raisebox{4pt}{\asymm}}\\
U_R&  \fund^m & 1^m & \frac{2}{3}  & \frac{1}{2}   \vphantom{\raisebox{4pt}{\asymm}}\\
D_R & \fund^m & 1^m & -\frac{1}{3}  & \frac{1}{2}  \vphantom{\raisebox{4pt}{\asymm}}\\
N_R & 1 & 1^m & 0    & -\frac{3}{2}     \vphantom{\raisebox{4pt}{\asymm}}\\
E_R & 1&1^m &  -1  & -\frac{3}{2}       \vphantom{\raisebox{4pt}{\asymm}}\\
\end{array}
\label{model2}
\eeq
where the superscript $m$ reminds us that there is a corresponding non-Abelian magnetic charge $\beta_a\ne0$. Note, that this is almost the same as in the toy model of section \ref{sec:Toymodel} but with smaller magnetic charges and additional non-Abelian magnetic charges.
In terms of the unbroken gauge symmetries we have
\beq
\begin{array}{c|ccc}
& SU(3)_c & U(1)_{em}^{el}& U(1)_{em}^{mag} \\
\hline
U_L & \fund^m & \frac{2}{3} &\frac{1}{2}   \vphantom{\raisebox{4pt}{\asymm}}\\
D_L & \fund^m & -\frac{1}{3} &\frac{1}{2}     \vphantom{\raisebox{4pt}{\asymm}}\\
N_L & 1 & 0 & -\frac{3}{2}    \vphantom{\raisebox{4pt}{\asymm}}\\
E_L & 1 & -1  &-\frac{3}{2}     \vphantom{\raisebox{4pt}{\asymm}}\\
U_R&  \fund^m & \frac{2}{3} &\frac{1}{2}      \vphantom{\raisebox{4pt}{\asymm}}\\
D_R & \fund^m & -\frac{1}{3} &\frac{1}{2}      \vphantom{\raisebox{4pt}{\asymm}}\\
N_R & 1 &0 &-\frac{3}{2}        \vphantom{\raisebox{4pt}{\asymm}}\\
E_R & 1& -1 &-\frac{3}{2}    \vphantom{\raisebox{4pt}{\asymm}}\\
\end{array}
\label{model2belowew}
\eeq
If we choose $\beta_c=1$ (in the normalization where $T^8=$diag(1/3,1/3,-2/3))
the Dirac-Schwinger quantization conditions are satisfied for all the fields in the model. Furthermore, the color magnetic charge contributes $T^8 g$ to the angular momentum of the gauge fields.
Finally, all anomalies are still canceled: the magnetic charges are still proportional to B-L and thus any combination of anomalies still has to vanish.

One immediate advantage of the model with smaller magnetic charges is that the Dirac quantization condition now allows smaller magnetic coupling:
\beq
\alpha^{mag}= \frac{\alpha^{-1}}{4}\sim 32 \ ,
\eeq
still a factor of few bigger than $4\pi$, but significantly smaller than in the previous case.

But even more important is the fact that Rubakov-Callan operator related to the top mass is generated in this model. Let us again consider scattering of a RH up-type quark on $N_L$. The angular momentum of the electromagnetic field is $\frac{2}{3}\times \frac{-3}{2}=-1$, while the spin of the incoming particles is +1. After the particles scatter in the forward direction, the angular momentum of the field flips, and that can be compensated by the simultaneous chirality flip of the top quark and the monopole. Thus an operator
\beq
\lambda^{(u)}_{ij} u_R^i N_L \left(u_L^j N_R \right)^\dagger
\eeq
should be present,  in fact we must have the full gauge invariant operator
\beq
\lambda^{(u)}_{ij} u_R^i L_L \left(q_L^j N_R \right)^\dagger~,
\eeq
which, after monopole condensation, can give rise to the large top mass, depending on the details of the flavor physics contained in $\lambda^{(u)}_{ij}$. In fact, a large mass for at least one of the up-type quarks is required by the consistency of the theory, whereas most extensions of the Standard Model would still be self-consistent with a 10 GeV top quark. Why there is only one heavy up-type quark is not explained. Indeed, monopole interactions can not break anomaly-free flavor symmetries, thus the appearance of a single heavy mass state requires an existence of a non-trivial flavor physics in the underlying UV theory. This UV physics has to break all non-anomalous flavor symmetries of the SM. In usual technicolor models the effects of such high-scale flavor violations would be strongly suppressed at low energies. Here however we can use the RC operators to transmit the high scale flavor violation to low scales without the strong suppression, thereby decoupling the scale of flavor physics from that of EWSB. Operators with four ordinary quarks do not involve strong magnetic charges, so they are suppressed by the UV scale of flavor physics, thus FCNC's could potentially be suppressed while keeping a heavy top mass. The detailed discussion of the UV flavor physics is beyond the scope of the current paper, however a  simple way to ensure unitarity in all possible scattering channels is for every element of $\lambda^{(u)}_{ij}$ to have approximately the same value, which (since it is approximately a rank one matrix) would only give one heavy mass, similar to what we observe in the real world.

What about the other quarks and leptons?  There are no four-fermion Rubakov-Callan operators that can generate the masses for the down-type quarks or leptons.  There are however six-fermion Rubakov-Callan operators that can generate these masses, however since they involve more ordinary fermions, the masses generated are suppressed by a loop-factor.    We find the following Rubakov-Callan scattering process:
\beq
d_R+E_L+u_L+d_L^\dagger \rightarrow u_L+E_R~,
\eeq
is allowed.
A down-type mass is then generated from this operator by closing up the up-quark loop. While this diagram appears naively to be quadratically divergent, the Rubakov-Callan operator will not be couple for very energetic external legs, which will provide the appropriate cutoff for these diagrams. Then we get a simple  estimate for the heaviest down-type quark mass of
\beq
m_b \sim \frac{m_t}{16 \pi^2}~,
\eeq
which is just a one-loop suppression relative to the top mass.  There are similar operators for the charged leptons.
Since neutrinos are electrically neutral, they cannot get Dirac masses through Rubakov-Callan operators in this type of model.  Thus the model necessarily requires that neutrinos are much lighter, just because they are neutral.  Dirac masses for the neutrinos can be radiatively generated from other fermion masses, this happens, for example, in Pati-Salam models
\cite{Appelquist} where the quark masses feed down to the neutrino masses.

Another generic problem of technicolor theories is an abundance of light pseudo-Nambu-Goldstone bosons.
In order to find the correct global symmetry of the magnetic sector we need to take into account additional RC operators
that are induced between the dyons, which can reduce the global symmetry.\footnote{ Just as in the case of the top mass the more precise statements is RC operators can transmit high scale flavor violation to low energies without a strong suppression by the high scale.}
The following four-fermion RC operators are induced between the dyons:
\beq
&& D_L E_R \left( E_L D_R \right)^\dagger\\
&&  U_L N_R  \left( N_L U_R \right)^\dagger~,
\label{RC}
\eeq
as well as their hermitian conjugates.
These two operators are contained in the gauge invariant operators
\beq
&& Q_L E_R \left( L_L D_R \right)^\dagger\\
&&  Q_L N_R  \left( L_L U_R \right)^\dagger~.
\label{gi}
\eeq
Assuming that the coefficients are equal, these operators reduce the global symmetry to
$SU(3)_c \times SU(2)_L  \times SU(2)_R  \times U(1)_Q \times U(1)_L$. The presence of these operators will ensure that there are no light pseudo-Nambu-Goldstone bosons in this model.

\section{Basic phenomenology}
\label{sec:pheno}
\setcounter{equation}{0}
\setcounter{footnote}{0}

Our model contains massless chiral monopoles which become vector-like after electroweak symmetry breaking (bilinear monopole condensation). It is then reasonable to expect that these monopoles themselves pick up a mass of the order of the monopole condensate $\Lambda_{mag} \sim$ 500 GeV - 1 TeV. One important question is whether these monopoles would be confined or not. As we have explained above that question crucially depends on the direction of magnetic charges in the space of gauge generators. In the model of Sec.~\ref{sec:model}, the alignment of the magnetic charges is chosen such that the magnetic field is a combination of the ordinary QED photon and the QCD gluon, both of which remain massless. Thus the electric group remains unbroken, implying that the monopoles will be unconfined. Of course the color magnetic charge will be screened at distances corresponding to $\Lambda_{QCD}$, so at long distances our monopoles would act as ordinary massive dyons of QED (some with minimal Dirac magnetic charge and some with three times this charge), subject to all of the collider and direct monopole searches applicable to them.
An advantage of this type of model is that the monopoles  have no magnetic coupling to the $Z$ boson, otherwise it would have been hard to imagine any possible way to avoid disastrously large corrections to electroweak precision measurements. Thus the electroweak precision operators are shielded from direct corrections from the strong magnetic interactions. However, there will still be corrections to $S$ and $T$ since the monopoles also carry electric charges (otherwise they couldn't possibly break the electroweak gauge symmetry). However, from the point of view of electroweak precision corrections the situation  will be similar to a fourth generation model, which can be made to agree with electroweak precision data if the masses within the fourth generation have the right splittings~\cite{Graham,Holdom4G}.

Another important question is the issue of CP violation in this model. As reviewed recently in~\cite{cst} theories with dyons generically do not conserve CP, unless a pairing of fields with charges $(q,g)$ and $(q,-g)$ is possible, which is not the case in our chiral model. So the main question is how these new sources of CP violation will feed down into the effective SM below the scale of monopole condensation.
There are two types of effects that one should consider. First, it is possible that the Rubakov-Callan operators of the type leading to fermion masses could also contribute to the CP violating observables, for example in $K-\bar K$ systems. The contributions of these operators, however, depends on the details of the flavor physics in the UV and they could be easily suppressed.
Second,
new CP violating electroweak precision operators of the form $W^3_{\mu\nu} B_{\alpha\beta}\,\epsilon^{\mu\nu\alpha\beta}$, i.e. a CP violating $S$ parameter, could be induced. However, these effects vanish in the limit when the SM fermions are massless, since in that case chiral rotations can eliminate these additional $\theta$-type terms from the effective Lagrangian. Thus such new electroweak CP violating effects should be proportional to the determinant of the mass matrix, including all fields charged under the $U(1)$ groups, which will make it proportional to neutrino masses and thus unobservably small given the current experimental precision. On the other hand new sources of CP violation may be useful in explaining the matter-antimatter asymmetry of the Universe~\cite{Holdom4G}.

Next we will comment briefly on the expected LHC phenomenology of this class of models, but leave a detailed analysis for later.  We expect that if the monopoles are light enough they could be pair produced at the LHC. Since their production is unsuppressed by gauge couplings the production cross section could be at most of the order of $4\pi/({\rm TeV})^2 \sim $ 1 nb. A more detailed simulation~\cite{Andi} shows that the actual expected cross sections at the LHC would be somewhat smaller, of the order of 1 pb. However since the Coulomb potential between a monopole anti-monopole pair is so strong they will likely radiate away any excess kinetic energy and annihilate into numerous photons. This is an interesting signal for experimentalists to look for, and indeed they are already looking for multi-photon events in the context of squirks \cite{squirks}. For monopole annihilation however we expect that there would be some harder photons than in the squirk case. The phenomenology of this model seems to be somewhat similar to that of Holdom's fourth generation model with a strongly coupled  $U(1)$ \cite{Holdom4G,Holdom}; one could produce various pairs of the fourth generation and with a collider of sufficient energy to produce free monopoles they could possibly undergo weak decays, depending on the exact mass spectrum. The main difference would be that the lightest particle with a magnetic charge must be absolutely stable.

One might wonder whether there are strong bounds on TeV mass monopoles from cosmic ray production.  The strongest bound on the monopole flux for masses below $5 \times 10^{13}$ GeV comes from the SLIM experiment \cite{SLIM} who find an upper bound of $1.3 \times 10^{-15}$ cm$^{-2}$ sr$^{-1}$ s$^{-1}$. In order to have a 10 TeV center of mass energy when scattering off a proton, a cosmic ray needs an energy of $10^8$ GeV, and the flux of such particles is around $10^{-22} $ cm$^{-2}$ sr$^{-1}$ s$^{-1}$ GeV$^{-1}$, and falling faster than the third power of energy. This can be used to put an upper bound on the monopole production cross section. Since the cosmic ray flux is falling very quickly with energy, the total flux above the threshold energy for monopole production can be estimated to be $10^{-14} $ cm$^{-2}$ sr$^{-1}$ s$^{-1}$. To obtain the secondary monopole flux from this, we can multiply by the proton density of the atmosphere/unit surface area of the Earth and the production cross section of the monopoles. This results in a conservative upper bound on the monopole cross section of order 0.1 mb, much above the expected pb size, implying that
cosmic ray searches for monopoles do not provide a strong bound on TeV scale monopoles.

Finally we note that in the absence of weak interactions  monopole number is a conserved quantity, and so including the effects of weak instantons there are two linear combinations of baryon, lepton, and monopole number that are conserved.
The lightest state carrying monopole number must be stable. However, due to the strong magnetic interactions, we expect relic monopoles to form magnetically (and also color) neutral bound states. These could either be  ``mesonic" (which can decay) and ``baryonic" composites of the monopoles, thus the lightest ``baryonic state" must be stable. If the lightest ``baryon" was also electrically neutral, like UDDN or UUDE, then this could be a dark matter candidate.  Whether such a heavy dark matter particle is allowed by direct detection experiments depends sensitively on the charge radius of this ``baryon".  This is not a question that can be answered by relativistic quantum mechanics, it would require a full lattice simulation with light quarks and leptons included.

\section{Conclusions}
\label{sec:conclude}
\setcounter{equation}{0}
\setcounter{footnote}{0}

In this paper we have presented an interesting new alternative for dynamical electroweak symmetry breaking based on a model with bilinear condensation of fermionic monopoles.  Perhaps its most interesting feature is that Rubakov-Callan operators could explain why the top quark is so heavy.  The top quark mass is usually the most difficult problem to solve in a model where strong dynamics breaks electroweak symmetry, but here no additional interactions are needed, the Rubakov-Callan operators are required for the consistency of a monopole theory. Whether the required operators exist in the theory depends simply on on the charge assignments of the fields.  Such models should have a variety of interesting signatures that can be searched for at the LHC and elsewhere.  Monopole annihilation into multi-photon final states with roughly a pb cross section would be relatively easy to find at the LHC with one fb$^{-1}$ of data since there is no standard model background.

\section*{Acknowledgements}
We thank   Louis Alvarez-Gaume, Sekhar Chivukula, Andy Cohen, Jacques Distler, Dan Green, Howie Haber, Markus Luty, Yaron Oz, Martin Schmaltz, Jay Wacker, and Mithat Unsal for useful
discussions and comments.  We also thank the Aspen Center for Physics, and JT thanks CERN, where part of this work was completed.  This research of C.C. has been supported in part by the NSF grant PHY-0757868 and in part by a U.S.-Israeli BSF grant. Y.S. is supported in part by the NSF grant  PHY-0653656.
J.T. is supported by the US Department of Energy grant DE-FG02-91ER40674.

\end{document}